# RankMat : Matrix Factorization with Calibrated Distributed Embedding and Fairness Enhancement

Hao, Wang*

Ratidar.com

Haow85@live.com

Beijing, China

Matrix Factorization is a widely adopted technique in the field of recommender system. Matrix Factorization techniques range from SVD, LDA, pLSA, SVD++, MatRec, Zipf Matrix Factorization and Item2Vec. In recent years, distributed word embeddings have inspired innovation in the area of recommender systems. Word2vec and GloVe have been especially emphasized in many industrial application scenario such as Xiaomi's recommender system. In this paper, we propose a new matrix factorization inspired by the theory of power law and GloVe. Instead of the exponential nature of GloVe model, we take advantage of Pareto Distribution to model our loss function. Our method is explainable in theory and easy-to-implement in practice. In the experiment section, we prove our approach is superior to vanilla matrix factorization technique and comparable with GloVe-based model in both accuracy and fairness metrics.

## 1  Introduction

Matrix factorization is a famous recommender system algorithmic framework. Although seemingly a bit out-of-dated, it is widely used even today with new breakthroughs widely applied in industry. Matrix factorization techniques such as Alternating Least Squares [1] have been included in important algorithmic packages such as Spark MLLib. In early days, the intuition behind matrix factorization is pretty simple, namely decreasing the RMSE or MAE. As the technology evolves, the optimization goals of the algorithm has diverged into different topics such as fairness. The merits of matrix factorization is very obvious : It's very easy to understand; it's very simple to implement; the speed and performance of the algorithm is acceptable in commercial environments.

In recent years, researchers and industrial workers have borrowed the idea of distributed word embeddings such as Word2Vec [2] and GloVe [3] to improve the performance of matrix algorithm. However, the official explanation of word embedding algorithms is usually obscure and hard to understand. The difficulty of the theory has not effected the application of the idea in the industry. Public materials have shown large companies such as Xiaomi have used embedding techniques to improve the user experiences of their products.

In this paper, we propose a new matrix factorization model inspired by and similar to GloVe [3] in appearance, but is entirely different in theory behind the technique. We illustrate the theory in the following sections of the paper and proves the superiority of our algorithm in the Experiment section.

---

* Place the footnote text for the author (if applicable) here.

## 2   Related Work

Matrix factorization is a commonly encountered approach in recommender system products. The basic idea behind the framework is pretty straightforward and easy to understand. The reduction in space complexity comes together with nice performance in time complexity and technical metrics. Popular matrix factorization approaches include SVD Feature [4], Alternating Least Squares [1], SVD++ [5], among a whole spectrum of machine learning techniques.

Researchers have kept keen interest in the possibility of borrowing ideas from other fields to enhance recommender system performance. Natural language processing inventions such as Latent Dirichlet Allocation [6] and Word2Vec [2] have been particularly emphasized in the field. Embedding techniques such as item2vec [7] have emerged as a standard recommender system technique in major technical companies.

Word2Vec [2] and GloVe [3] are the 2 most popular word distributed representation technologies in the natural language processing field. The basic intuition behind the techniques is to seek dense representation of words in relation of probabilistic formulas of co-occurence probability and conditional probability. The initial explanation of the theory behind these 2 techniques are quite obscure and hard to understand. In this paper, we give a much simpler interpretation of the GloVe [3] model and come up with an entirely new formulation for recommender system domain.

In recent years, researchers have been more and more interested in the fairness problem of AI approaches. An ACM conference called FaCCT was established in aim to bring together world researchers to discuss fairness problems in the venue. A fairness-based matrix factorization called MatRec [8] was invented in 2020 incorporating rank information of users and items in the matrix factorization framework. One year later, H.Wang [9] invented a new algorithm called Zipf Matrix Factorization that enhances technical accuracy and fairness metrics at the same time. Fairness is also a hot topic in the area of Learning to Rank [10][11].

## 3   RankMat formulation

The initial formulation of matrix factorization is as follows : We decompose the user item rating matrix into the dot product of user feature vectors and item feature vectors to improve the performance of recommender system. In this way, the space complexity of the parameters is reduced to O(n), so matrix factorization technique can be viewed as a dimensionality reduction technique. The vanilla matrix factorization problem formulation is as follows :

$$RMSE = \sum_{i=1}^{m} \sum_{j=1}^{n} \left(u_i \bullet v_j - R_{ij}\right)^2$$

Modifications of the problem formulation leads to different matrix factorization. Some researchers have borrowed ideas from natural language processing and considered the user rating matrix as the word occurence matrix in document corpus. Such approaches include Latent Dirichelet Allocation based recommender system, and lately recommender systems that are inspired by word embedding technologies such as Word2Vec and GloVe. However, some techniques like item2vec do not factorize user-item rating matrix. Item2vec factorizes the item coocurrence matrix and uses the vectorized item data in the similarity computation procedure of collaborative filtering.

Unlike item2vec, we propose a new algorithm in this paper called RankMat that decomposes the user-item-rating matrix directly, instead of the item-item co-occurence matrix. The intuition behind our algorithm is as follows: Take the implicit feedback scenario as an example, if we fill the user-item rating matrix with the number of implicit feedbacks rather than the usual 0-1 values, we have a user-item cooccurrence matrix that fits the theory of word embeddings. For the explicit feeback, the rating the user gives to an item can be considered as the user-item cooccurence matrix : The rating values can be considered as the number of implicit feedbacks the user receives.

The formal problem formulation of GloVe is as follows :

$$L = \sum_{i,j=1}^{V} f(X_{ij}) \left(w_i^T \bullet w_j + b_i + b_j - \log X_{ij}\right)^2$$

Borrowing this idea into the recommender system, we obtain the following loss function for matrix factorization (taking out the bias term and weighting function) :

$$L = \sum_{i=1}^{m}\sum_{j=1}^{n}\left(u_i^T \bullet v_j - \log(R_{ij}+1)\right)^2$$

We call this algorithm GloVeMat. There have been many explanations behind the theory of GloVe, including the formal interpretation of the official paper. However, we consider GloVe as a simple model that models the inner product of embeddings as the power exponent of an exponential distribution that captures the distribution of word cooccurences as follows :

$$E = e^{w_i^T \bullet w_j + b_i + b_j} - X_{ij}$$

The initial formulation of GloVe is essentially the log of E applied in kernel smoothing setting. Using exponential function to model the distribution of word coocurrences is actually inaccurate, because as well known in Linguistics, distribution of words in corpuses follow Zipf's Law (Or more generally, Power Law Distribution), so we model the problem differently as follows:

$$E = \left(\frac{1}{rank_i} \times \frac{1}{rank_j}\right)^{u_i^T \bullet v_j} - X_{ij}$$

, which can be simplified in the following way (taking out the bias term) as the loss function of our matrix factorization approach:

$$L = \sum_{i=1}^{m}\sum_{j=1}^{n}\left(u_i^T \bullet v_j - \frac{\log(R_{ij}+1)}{\log(rank_i+1)+\log(rank_j+1)}\right)^2$$

We name our algorithm RankMat, and in the following section, we illustrate by experiments that our approach yields good performance compared with vanilla matrix factorization and GloVeMat, but at the same time is more robust to the Matthew Effect problem, as measured by the fairness metric proposed by H. Wang [9]. A side conclusion of our experiment is that GloVeMat itself is superior to vanilla matrix factorization algorithm solved by Stochastic Gradient Descent.

Please notice all the algorithms including vanilla matrix factorization, GloVeMat and RankMat are solved using Stochastic Gradient Algorithm in the next section. However, in practice, better optimization techniques could be used to solve the loss functions.

## 4 Experiment

We use a subset of MovieLens dataset that contains 610 users and 9742 movies to test vanilla matrix factorization, GloVeMat and RankMat. The evaluation metrics are MAE for technical accuracy and Degree of Matthew Effect [9] for fairness evaluation. All 3 algorithms tested in this section are resolved by Stochastic Gradient Descent. We compare the results of the 3 different algorithms by doing grid search on gradient learning step values :

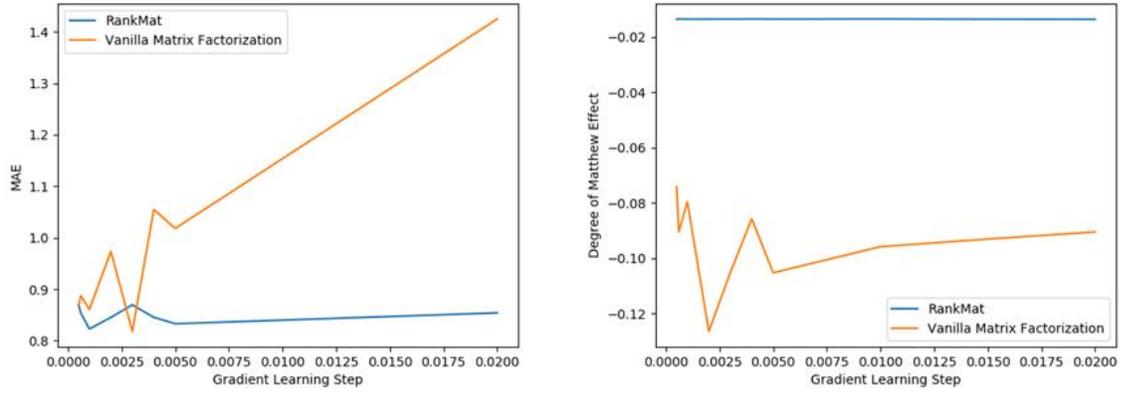

**Fig.1 Comparison** between Vanilla Matrix Factorization and RankMat in MAE (left) and Degree of Matthew Effect (Right)

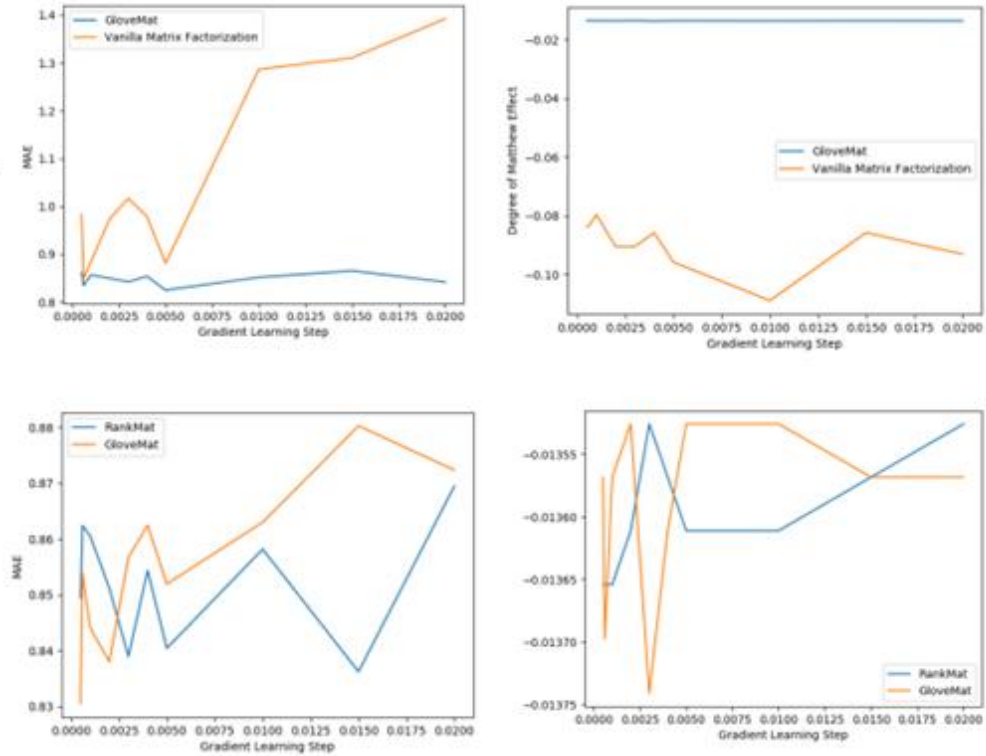

**Fig.2 Comparison** between Vanilla Matrix Factorization and GloVeMat in MAE (left) and Degree of Matthew Effect (Right)

Fig.2 Comparison between RankMat and GloVeMat in MAE (left) and Degree of Matthew Effect (Right)

From the figures illustrated above, we know that both GloVeMat and RankMat outperforms Vanilla Matrix Factorization in fairness metrics: Their Degree of Matthew Effect values are consistently larger than that of Vanilla Matrix Factorization. GloVeMat outperforms RankMat in fairness metric mostly. RankMat outperforms Vanilla Matrix Factorization in most grid search values of gradient descent learning step with an exception in a very small subinterval. GloVeMat outperforms Vanilla Matrix Factorization consistently in all grid search values. RankMat outperforms GloVeMat after a threshold and seems to be more robust because nice performance on small gradient learning steps might be due to precision error.

The experiments prove that GloVeMat and RankMat are both very good choices for matrix factorization model. GloVeMat has better fairness performance than RankMat while RankMat has better accuracy performance than GloVeMat. Both of them are supeior to the vanilla matrix factorization algorithm.

# 5 Conclusion

In this paper, we propose a new matrix factorization model called RankMat inspired by the GloVe model and Pareto Distribution. The method is theoretically rigorous and only needs a small modification to the original formulation of matrix factorization. Alongside with another algorithm proposed in this paper called GloVeMat, we prove the superiority of these techniques over the primitive matrix factorization formulation. In future work, we would like to explore other probabilistic modeling techniques for the matrix factorization framework and the application of other optimization solvers such as Adam and Adagrad.